\documentclass[11pt,titlepage,a4paper]{article}
\usepackage{bozhomac}	
\usepackage{bozhlogo}	
\usepackage{cite}

\usepackage{xr1}
\title{\bfseries	\vspace*{-2.199in}
{\huge Fibre bundle formulation of \\[0.22ex] relativistic quantum mechanics}
\\[1.1ex]
{\LARGE  II. Covariant approach}
}

\vspace{1.7ex}

\author{
Bozhidar Z. Iliev
\thanks{Department Mathematical Modeling,
Institute for Nuclear Research and \mbox{Nuclear} Energy,
Bulgarian Academy of Sciences,
Boul. Tzarigradsko chauss\'ee~72, 1784 Sofia, Bulgaria}
\thanks{E-mail address: bozho@inrne.bas.bg}
\thanks{URL: http://theo.inrne.bas.bg/$\sim$bozho/}
}

\date{
 \vspace{2.27ex}\ShortTitle{Bundle relativistic quantum mechanics: II}
								\\[0.27ex]
 \vspace{3.27ex}
\small
	\begin{tabular}{r@{$\colon\to~$}l}
 \vspace{0.09ex} Basic ideas	& November 1997, January 1998	\\[0.09ex]
 \vspace{0.09ex} Began/Ended	& February 16, 1998/March 27, 1998 \\[0.09ex]
 \vspace{0.09ex} Initial typeset& April 1--14, 1998	\\[0.09ex]
 \vspace{0.09ex} Revised	& August 1999, March 2002	\\[0.09ex]
  \vspace{0.09ex} Last update	& March 30, 2002		\\[1.09ex]
\vspace{0.27ex} Composing/Extracting part II
			& April 22/April 24, 1998	\\[0.27ex]
\vspace{0.27ex}  Updating part II& October, 1998, August 1999, March 2002
							\\[0.27ex]
  \vspace{0.27ex} Produced	& \fbox{\today}		\\[0.27ex]
	\end{tabular} \\[1.27ex]
\normalsize
	\begin{tabular}{r@{$\colon~$}l}
\vspace{0.27ex} LANL arXiv server E-print No. & quant-ph/0107002
 						\\[0.27ex]
	\end{tabular} \\[-0.27ex]
 \vspace{4.27ex}{\Huge\BOZHO}	\\[4.27ex]
 \vspace{0.27ex}\Subject{Relativistic quantum mechanics,
			  Differential geometry}		\\[2.27ex]
	\begin{tabular}{r@{\hspace{0.512em}}|@{\hspace{0.512em}}l}
 \vspace{0.27ex}		
&
 \vspace{0.27ex}
	\end{tabular} \\[1.27ex]
 \vspace{0.27ex}	\\[0.27ex]
}

\newcommand{\fibre}{\mathcal{F}} 
\newcommand{\bundle}{(\bspace,\pr,\basesp)}	
	\newcommand{\bspace}{\mathnormal{F}}	
	\renewcommand{\pr}{\pi}			
	\newcommand{\basesp}{\mathnormal{M}}	
\newcommand{\fibreover}[1]{{\bspace_{#1}}} 

	\newcommand{\bHil}{\mathit{F}}	

\newcommand{\Ham}{\mathcal{H}}	
\newcommand{\bHam}{\mathit{H}}	

	\newcommand{\ope}[1]{\mathcal{#1}}		 
	\newcommand{\mor}[1]{\mathit{#1}}		 

\begin{document}		

\renewcommand{\thefootnote}{\fnsymbol{footnote}} 
\maketitle				
\renewcommand{\thefootnote}{\arabic{footnote}}   

\tableofcontents		


\begin{abstract}

	We propose a fibre bundle formulation of the mathematical base of
relativistic quantum mechanics. At the present stage the bundle form of the
theory is equivalent to its conventional one, but it admits new types
of generalizations in different directions.

	In the present, second, part of our investigation, we consider a
covariant approach to bundle description of relativistic quantum mechanics.
	In it the wavefunctions are replaced with (state) sections of a
suitably chosen vector bundle over space-time whose (standard) fibre is the
space of the wavefunctions. Now the quantum evolution is described as a
linear transportation (by means of the transport along the identity
map of the space\nobreakdash-time) of the state sections in the (total)
bundle space. Connections between the (retarded) Green functions of the
relativistic wave equations and the evolution operators and transports are
found. Especially the Dirac and Klein-Gordon equations are considered.

\end{abstract}

\section {Introduction}
\label{II.Introduction}

	This paper is a second part of our investigation devoted to the fibre
bundle description of relativistic quantum mechanics. It is a
straightforward continuation of~\cite{bp-BRQM-time-dependent}.

	The developed in~\cite{bp-BRQM-time-dependent} bundle formalism for
relativistic quantum wave equations has the deficiency that it is not
explicitly covariant; so it is not in harmony with the relativistic theory it
represents. This is a consequence of the direct applications of the bundle
methods developed for the nonrelativistic region, where they work well
enough, to the relativistic one. The present paper is intended to mend this
`defect'. Here we develop an appropriate covariant bundle description of
relativistic quantum mechanics which corresponds to the character of this
theory.

	The organization of the material is the following.

	Sect.~\ref{Sect7} contains a covariant application of the
ideas of the bundle description of nonrelativistic quantum mechanics
(see, e.g.~\cite[sect.~2]{bp-BRQM-time-dependent}
	or~\cite{bp-BQM-interpretation+discussion,bp-BQM-full})
to Dirac equation.  The bundle, where Dirac particles `live', is a vector
bundle over space-time with the space of 4-spinors as a fibre; so here we
work again with the 4-spinor bundle of~\cite[sect.~4]{bp-BRQM-time-dependent},
but now the evolution of a Dirac particle is described via
a \emph{ geometric transport} which is a
\emph{linear transport along the identity map of space-time}.
The state of a Dirac particle is represented by a
\emph{section (not along paths!)}
of the 4-spinor bundle and is (linearly) transported by means
of the transport mentioned. The Dirac equation itself is transformed into a
covariant Schr\"odinger-like equation.

	In Sect.~\ref{Sect8} we apply the covariant bundle approach to
Klein-Gordon equation. For this purpose we present a 5-dimensional
representation of this equation as a first-order Dirac-like equation to which
the theory of Sect.~\ref{Sect7} can be transferred \emph{mutatis mutandis}.

	The goal of Sect.~\ref{Sect9} is to be revealed some connections
between the retarded Green functions ($\equiv$propagators) of the
relativistic wave equations and the corresponding to them evolution operators
and transports.  Generally speaking, the evolution operators (resp.\
transports) admit representation as integral operators, the kernel of which
is connected in a simple manner with the retarded
Green function (resp.\ Green morphism of a bundle).
Subsect.~\ref{Subsect9.0} contains a brief general consideration of the Green
functions and their connection with the evolution transports, if any. In
Subsect.~\ref{Subsect9.1}, \ref{Subsect9.2}, and~\ref{Subsect9.3} we derive
the relations mentions for Schr\"odinger, Dirac, and Klein-Gordon equations,
respectively.

	Sect.~\ref{II.Conclusion} closes the paper with a brief summary of
the main ideas underlying the bundle description of relativistic quantum
mechanics.

	\ref{AppendixA} contains some mathematical results
concerning the theory of (linear) transports along maps required for the
present investigation.

	In~\ref{AppendixB} are given certain formulae concerning matrix
operators, \ie matrices with operator entries, which arise naturally in
relativistic quantum mechanics.
\vspace{1.2ex}

	The notation of the present work is the the same as the one
in~\cite{bp-BRQM-time-dependent}
and we are not going to recall it here.

	The references to sections, equations, footnotes etc.
from~\cite{bp-BRQM-time-dependent} are obtained from their sequential
numbers in~\cite{bp-BRQM-time-dependent} by adding in front of them the
Roman one (I) and a dot as a separator. For instance, Sect.~I.4 and (I.5.2)
mean respectively section 4 and equation~(5.2) (equation 2 in Sect.~5)
of~\cite{bp-BRQM-time-dependent}.

	Below, for reference purposes, we present a list of some essential
equations of~\cite{bp-BRQM-time-dependent} which are used in this paper.
Following the just given convention, we retain their original reference
numbers.
%
	\begin{gather}
\tag{\ref{2.2}}
\psi(t) = \ope{U}(t,t_0)\psi(t_0),
\\
\tag{\ref{2.3}}
\Psi_\gamma\colon t\to \Psi_\gamma(t) = l_{\gamma(t)}^{-1}\bigl(\psi(t)\bigr),
\qquad
\bHam_\gamma\colon t\to
\bHam_\gamma(t) = l_{\gamma(t)}^{-1} \circ \Ham \circ l_{\gamma(t)},
\displaybreak[1]\\
\tag{\ref{2.4}}
\mor{U}_\gamma(t,s)
=l_{\gamma(t)}^{-1}\circ \ope{U}(t,s) \circ l_{\gamma(s)}
\colon\bHil_{\gamma(s)}\to\bHil_{\gamma(t)},
\qquad s,t\in J,
\displaybreak[1]\\
\tag{\ref{2.11}}
\mor{A}_{\gamma}(t) =
l_{\gamma(t)}^{-1} \circ \ope{A}(t) \circ l_{\gamma(t)}
\colon  \bHil_{\gamma(t)}\to\bHil_{\gamma(t)}.
	\end{gather}

\section {Dirac equation}
\label{Sect7}

	The covariant Dirac
equation~\cite[sect.~2.1.2]{Itzykson&Zuber},
\cite[chapter~XX, \S~8]{Messiah-2}
for a (spin $\frac{1}{2}$) particle with mass $m$ and electric charge $e$ in
an external electromagnetic field with 4\ndash potential $\ope{A}^\mu$ is
	\begin{equation}	\label{7.1}
(\ih\Slashed{D} - mc\openone_4) \psi = 0,
\qquad
\Slashed{D}:=\gamma^\mu D_\mu, \quad
			D_\mu:=\pd_\mu -\frac{e}{\ih c}\ope{A}_\mu .
	\end{equation}
Here $i\in\mathbb{C}$ is the imaginary unit, $\hbar$ is the Planck constant
(divided by $2\pi$),
$\openone_4=\diag(1,1,1,1)$ is the $4\times4$ unit matrix,
$\psi:=(\psi^0,\psi^1,\psi^2,\psi^3)$ is (the matrix of the components of) a
4\ndash spinor, $\gamma^\mu$, $\mu=0,1,2,3$, are the well known Dirac
$\gamma$\ndash matrices~\cite{Bjorken&Drell-1, Messiah-2,Itzykson&Zuber}, and
$c$ is the velocity of light in vacuum.
	Since~\eref{7.1} is a first order partial differential equation on
the Minkowski 4\ndash dimensional spacetime, it does not admit an evolution
operator with respect to the spacetime. More precisely, if $x_1,x_2\in M$,
$M$ being the spacetime, \ie the Minkowski space $M^4$, then there does
\emph{not} exist a $4\times4$ matrix operator (see~\ref{AppendixB})
$\ope{U}(x_2,x_1)$ such that
	\begin{gather*}	
\psi(x_2) = \ope{U}(x_2,x_1) \psi(x_1).
\\
(\ih\Slashed{D}_x - mc\openone_4) \ope{U}(x,x_0) = 0, \qquad
\ope{U}(x_0,x_0) = \id_\fibre, \quad x,x_0\in M
	\end{gather*}
where
$\Slashed{D}=\slashed{\pd} - \frac{e}{\ih c} \Slashed{\ope{A}}$
with
$\slashed{\pd}:=\gamma^\mu\pd/\pd x^\mu$ and
$\Slashed{\ope{A}}:=\gamma^\mu\ope{A}_\mu$,
$\fibre$ is the space of 4\ndash spinors,
and $\id_X$ is the identity map of a set $X$.

	By this reason, the methods of~\cite{bp-BRQM-time-dependent} cannot
be applied directly to the general 4-dimensional spacetime descriptuion of
the Dirac equation. An altternative approach to the problem is presented
below.

	Suppose $\bundle$ is a vector bundle with (total) bundle space
$\bspace$, projection $\pr\colon\bspace\to\basesp$, fibre $\fibre$,
and isomorphic fibres $\fibreover{x}:=\pr^{-1}(x)$, $x\in\basesp$.
There exist linear isomorphisms $l_x:\fibreover{x}\to\fibre$ which
we assume to be diffeomorphisms; so
$\fibreover{x}=l_{x}^{-1}(\fibre)$ are 4\ndash dimensional vector spaces.

	To a state vector (spinor) $\psi(x_0)$ at a fixed point $x_0$, we
assign a $C^1$ section%
\footnote{%
In contrast to the time\ndash dependent approach~\cite{bp-BRQM-time-dependent}
and nonrelativistic case~\cite{bp-BQM-introduction+transport}
now $\Psi_{x_0}$ is simply a section, not section along
paths~\cite{bp-BQM-introduction+transport}.  Physically this corresponds to
the fact that quantum objects do not have world lines (trajectories) in a
classical sense~\cite{Messiah-1}.%
}
$\Psi_{x_0}$ of $\bundle$, i.e.\ $\Psi_{x_0}\in\Sec^1\bundle$, by
(cf.~\eref{2.3})
	\begin{equation}	\label{7.4}
\Psi_{x_0}(x)
:= l_{x}^{-1}\bigl(\psi(x_0)\bigr) \in\fibreover{x}:=\pr^{-1}(x),
\quad x\in M.
	\end{equation}
Generally $\Psi_{x_0}(x)$ depends on the choice of the point $x_0\in\basesp$.

	Since in $\bundle$ the state of a Dirac particle is described by
$\Psi_{x_0}$, we call it \emph{state section}; resp.\ $\bundle$ is the
\emph{4\ndash spinor bundle}.
The description of Dirac particle
via $\Psi_{x_0}$ will be called \emph{bundle description}. If it is known, the
conventional spinor description is achieved by the spinor
	\begin{equation}	\label{7.4'}
\psi(x_0) := l_{x}\bigl(\Psi_{x_0}(x)\bigr) \in\fibre.
	\end{equation}
Evidently, we have
	\begin{equation}	\label{7.5}
\Psi_{x_0}(x_2) = \mor{L}(x_2,x_1) \Psi_{x_0}(x_1),\qquad x_2,x_1\in M.
	\end{equation}
with (cf.~\eref{2.4})
	\begin{equation}	\label{7.6}
\mor{L}(y,x) :=
l_{y}^{-1} \circ l_{x}\colon\fibreover{x}\to\fibreover{y},
\qquad x,y\in M.
	\end{equation}

	Obviously, $\ope{L}$ is a linear $4\times4$ matrix operator
satisfying the equations
	\begin{align}	\label{7.8}
\mor{L}(x_3,x_1) &= \mor{L}(x_3,x_2) \circ \mor{L}(x_2,x_1),
\qquad x_1,x_2,x_3\in M,
\\
	\label{7.9}
\mor{L}(x,x) &= \id_{\fibreover{x}}, \qquad x\in M.
	\end{align}

	Consequently, by definition~\ref{DefnLT.1}, the map
$\mor{L}\colon(y,x)\to\mor{L}(y,x)=\mor{K}_{x\to y}^{\id_M}$ is a linear
transport along the identity map $\id_M$ of $M$ in the bundle $\bundle$.
Alternatively, as it is mentioned in~\ref{AppendixA}, this means that
\(
\mor{L}^{\gamma}\colon(t,s) \to
\mor{L}_{s\to t}^{\gamma}:=\mor{L}(\gamma(t),\gamma(s)),\
s,t\in J
\)
is a flat linear transport along $\gamma\colon J\to M$ in $\bundle$.
(Besides, $\mor{L}(y,x)$ is a Hermitian and unitary transport ---
see~\cite{bp-BQM-introduction+transport}.)

	Equation~\eref{7.5} simply means that $\Psi_{x_0}$ is
$\mor{L}$\ndash\emph{transported (along $\id_M$) section} of $\bundle$
(cf.~\cite[definition~5.1]{bp-LTP-appl}). Writing~\eref{LT.10} for the
transport $\mor{L}$ and applying the result to a state section
given by~\eref{7.4}, one can prove that~\eref{7.5} is equivalent to
	\begin{equation}	\label{7.10}
\mathcal{D}_\mu \Psi = 0, \qquad \mu=0,1,2,3
	\end{equation}
where, for brevity, we have put
$\mathcal{D}_\mu:=\mathcal{D}_{x^\mu}^{\id_M}$ which is the
$\mu$\nobreakdash-th partial (section\ndash)derivation along the identity map
(of the spacetime) assigned to the transport $\mor{L}$.

	Now we shall introduce local bases and take a local view of the
above\ndash described material.

	Let $\{f_\mu(x)\}$ be a basis in $\fibre$ and $\{e_\mu(x)\}$ be a
basis in $\fibreover{x},\ x\in M$.
The matrices corresponding to vectors and/or
linear maps (operators) in these fields of bases will be denoted by the same
(kernel) symbol but in \textbf{boldface}, for instance:
$\Mat{\psi}:=\bigl(\psi^0,\psi^1,\psi^2,\psi^3\bigr)^\top$ and
$\Mat{l}_{x}(y):=\bigl[(l_x)_{~\nu}^{\mu}(y)\bigr]$
are defined, respectively, by
 $\psi(x)=:\psi^\mu(x)f_\mu(x)$ and
$l_x\bigl(e_\nu(x)\bigr)=:\bigl(l_x(y)\bigr)_{~\nu}^{\mu}f_\mu(y)$.
We put $\Mat{l}_x:=\Mat{l}_x(x)$; in fact this will be the only case when the
matrix of $l_x$ will be required as we want the `physics in
$\fibreover{x}$' to
correspond to that of $\fibre$ at $x\in M$. A very convenient choice is to
put $e_\mu(x)=l_{x}^{-1}\bigl(f_\mu(x)\bigr)$; so then
$\Mat{l}_x=\openone_4:=\bigl[\delta_{\nu}^{\mu}\bigr]=\diag(1,1,1,1)$ is the
$4\times4$ unit matrix.

	The matrix elements of the mapping $\mor{L}(y,x)$ are defined via the
equation
$\mor{L}(y,x)\bigl(e_\mu(x)\bigr)=:\mor{L}_{~\mu}^{\lambda}(y,x)e_\lambda(y)$
and, due to~\eref{7.6}, we have
	\begin{equation}	\label{7.12}
\Mat{\mor{L}}(y,x) =
	\Mat{l}_{y}^{-1} \cdot \Mat{l}_{x}
	\end{equation}
which is generically a matrix operator (see~\ref{AppendixB}).

	According to~\eref{LT.12} the \emph{coefficients} of the
transport $\mor{L}$ form four matrix operators
	\begin{equation}	\label{7.13}
\lindex[\Mat{\Gamma}]{\mu}{}(x) :=
\bigl[ \lindex[\Gamma]{\mu}{} _{~\nu}^{\lambda}(x) \bigr]_{\lambda,\nu=0}^{3}
  := \left. \frac{\pd\Mat{\mor{L}}(x,y)}{\pd y^\mu}\right|_{y=x}
   = \Mat{l}^{-1}(x) \frac{\pd\Mat{l}(x)}{\pd x^\mu}
	\end{equation}
where~\eref{7.12} was applied (see also theorem~\ref{ThmLT.1}).

	Applying~\eref{LT.5} and~\eref{7.13}, we find
	\begin{equation}	\label{7.14}
\frac{\pd\Mat{\mor{L}}(y,x)}{\pd y^\mu} =
- \lindex[\Mat{\Gamma}]{\mu}{}(y) \odot \Mat{\mor{L}}(y,x), \qquad
\frac{\pd\Mat{\mor{L}}(y,x)}{\pd x^\mu} =
\Mat{\mor{L}}(y,x) \odot \lindex[\Mat{\Gamma}]{\mu}{}(x) ,
	\end{equation}
where $\odot$ denotes the introduced by~\eref{mo.2} multiplication of matrix
operators. Therefore
	\begin{equation}	\label{7.15}
\slashed{\pd}_y \Mat{\mor{L}}(y,x)  =
	- \Slashed{\Mat{\Gamma}}(y) \Mat{\mor{L}}(y,x),
\qquad
\Slashed{\Mat{\Gamma}}(x) :=
	\gamma^\mu \cdot \lindex[\Mat{\Gamma}]{\mu}{}(x) .
	\end{equation}

	Similarly to the nonrelativistic
case~\cite{bp-BQM-equations+observables}, to any operator
$\ope{A}\colon\fibre\to\fibre$ we assign a bundle morphism
$\mor{A}\colon\bspace\to\bspace$ by
	\begin{equation}	\label{7.16}
\mor{A}_x := \mor{A}|_{\fibreover{x}} := l_{x}^{-1}\circ\ope{A}\circ l_x.
	\end{equation}
Defining
	\begin{equation}	\label{7.17}
\mor{G}^\mu(x) := l_{x}^{-1}\circ\gamma^\mu\circ l_x,
\quad
d_\mu := d_\mu|_x :=  l_{x}^{-1}\circ\pd_\mu\circ l_x,
\qquad
\pd_\mu:=\frac{\pd}{\pd x^\mu}
	\end{equation}
and using the matrices $\ope{G}^\mu(x)$ and $\ope{E}_\mu(x)$ given
via~\eref{mo.11}, we get
	\begin{equation}	\label{7.18}
\Mat{\mor{G}}^\mu(x) = \Mat{l}_{x}^{-1}\ope{G}^\mu(x) \Mat{l}_x,
\quad
\Mat{d}_\mu
:= \openone_4\pd_\mu +
   \Mat{l}_{x}^{-1}\bigl(\pd_\mu\Mat{l}_x + \ope{E}_\mu(x)\Mat{l}_x \bigr).
	\end{equation}
(Here and below, for the sake of shortness, we sometimes omit the argument
$x$.)
	The anticommutation relations
	\begin{equation}	\label{7.19}
	\begin{split}
\mor{G}^\mu\mor{G}^\nu + \mor{G}^\nu\mor{G}^\mu
& = 2\eta^{\mu\nu}\id_\bspace,
\\ 
\Mat{\mor{G}}^\mu\Mat{\mor{G}}^\nu + \Mat{\mor{G}}^\nu\Mat{\mor{G}}^\mu
& = \ope{G}^\mu\ope{G}^\nu + \ope{G}^\nu\ope{G}^\mu
= 2\eta^{\mu\nu}\openone_4
\quad
\bigl(= \gamma^\mu\gamma^\nu + \gamma^\nu\gamma^\mu \bigr) ,
	\end{split}
	\end{equation}
where $[\eta^{\mu\nu}]=\diag(1,-1,-1,-1)=[\eta_{\mu\nu}]$ is the Minkowski
metric tensor, can be verified by means of~\eref{7.17}, \eref{7.18},
\eref{mo.9}, and the well know analogous relation for the
$\gamma$\ndash matrices (see, e.g.~\cite[chapter~2,
equation~(2.5)]{Bjorken&Drell-1}).

	For brevity, if $a_\mu\colon\bspace\to\bspace$ are morphisms, sums
like $\mor{G}^\mu\circ a_\mu$ will be denoted by `backslashing' the kernel
letter, $\backslashed{a}:=\mor{G}^\mu\circ a_\mu$ (\cf the `slashed' notation
$\slashed{a}:=\gamma^\mu a_\mu$). Similarly, we put
$\backslashed{\Mat{a}}:=\Mat{\mor{G}}^\mu(x)\Mat{a}_\mu$ for
$\Mat{a}_\mu\in GL(4,\mathbb{C})$.
It is almost evident (see~\eref{7.17}) that the morphism corresponding to
$\slashed{\pd}:=\gamma^\mu\pd_\mu$ is
$\backslashed{d}=\mor{G}^\mu(x)\circ d_\mu$:
	\begin{equation}	\label{7.20}
\backslashed{d} = l_{x}^{-1} \circ \slashed{\pd} \circ l_x .
	\end{equation}

	Now we can write the Dirac equation~\eref{7.1} in a bundle form.

	First of all, we rewrite~\eref{7.1} as
	\begin{equation}	\label{7.22}
\ih\slashed{\pd}_x\psi(x) = \mathcal{D}_x\psi(x),
\qquad
\mathcal{D}_x := mc\openone_4 + \frac{e}{c}\Slashed{\ope{A}}(x) ,
	\end{equation}
the index $x$ meaning that the corresponding operators act with respect to
the variable $x\in\basesp$. This is the \emph{covariant}
Schr\"odinger\nobreakdash-like form of Dirac equation; $\slashed{\pd}_x$ is
the analogue of the time derivation $\od/\od t$ and $\mathcal{D}_x$
corresponds to the Hamiltonian $\Ham$. We call $\mathcal{D}$ the \emph{Dirac
function}, or simply, \emph{Diracian} of a particle described by Dirac
equation.

	Substituting~\eref{7.4'} into~\eref{7.22}, acting on the result from
the left by $l_{y}^{-1}$, and using~\eref{7.17}, we find the bundle form
of~\eref{7.22} as
	\begin{equation}	\label{7.23}
\ih\backslashed{d}_x|_y \Psi_x(y) = D_x|_y\Psi_x(y)
	\end{equation}
with $y\in\basesp$, $\backslashed{d}_x|_y:=\mor{G}(y) {d_\mu}|_x|_y$,
 ${d_\mu}|_x|_y:=l_y^{-1}\circ\frac{\pd}{\pd x^\mu}\circ l_y$
$D_x\in\Morf\bundle$ being the (Dirac) bundle morphism assigned to the
Diracian. We call it the \emph{bundle Diracian}. According to~\eref{7.16} it
is defined by
	\begin{equation}	\label{7.24}
D_x|_y := D_x|_{\fibreover{y}} = l_{y}^{-1}\circ\mathcal{D}_x\circ l_y
     = mc\id_{\fibreover{y}} + \frac{e}{c}\Backslashed{\mor{A}}_x|_y ,
	\end{equation}
where
 $\mor{A}_\mu|_x=\ope{A}_\mu|_x \id_\bspace\in\Morf\bundle$,
\(
\mor{A}_\mu|_x|_y
=
l_{y}^{-1}\circ(\ope{A}_\mu|_x\id_\fibre)\circ l_y
=
\ope{A}_\mu|_x\id_{\fibreover{y}}
\)
are the components of the \emph{bundle electromagnetic potential} and
	\begin{multline}	\label{7.25}
\Backslashed{\mor{A}}_x |_y
= \mor{G}^\mu(y)\circ\mor{A}_\mu|_x |_y
= \mor{G}^\mu(y)\circ (l_y^{-1}\circ \ope{A}_\mu\id_\fibre \circ l_y )
= l_{y}^{-1}\circ( \Slashed{\ope{A}}|_x\id_\fibre )\circ l_y
= \Backslashed{\ope{A}}_x |_y.
	\end{multline}

\section {Other relativistic wave equations}
\label{Sect8}

	The relativistic-covariant Klein-Gordon equation for a (spinless)
particle of mass $m$ and electric charge $e$ in a presence of (external)
electromagnetic field with 4\ndash potential $\ope{A}_\mu$
is~\cite[chapter~XX, \S~5, equation~(30$^\prime$)]{Messiah-2}
	\begin{equation}	\label{8.1}
\Bigl( \mathcal{D}^\mu\mathcal{D}_\mu +\frac{m^2c^2}{\hbar^2} \Bigr)\phi = 0,
\qquad
\mathcal{D}_\mu = \eta_{\mu\nu}\mathcal{D}^\nu
:=\pd_\mu - \frac{e}{\ih c}\ope{A}_\mu.
	\end{equation}
Since this is a second-order partial differential equation, it does not
directly admit an evolution operator and adequate bundle formulation and
interpretation. To obtain such a formulation, we have to rewrite~\eref{8.1}
as a first\ndash order (system of) partial differential equation(s) (\cf
Sect.~\ref{Sect5}).

	Perhaps the best way to do this is to replace $\phi$ with a
$5\times1$ matrix $\varphi=(\varphi^0,\ldots,\varphi^4)^\top$ and to
introduce $5\times5$ $\Gamma$\ndash matrices $\Gamma^\mu$,  $\mu=0,1,2,3$
with components $\bigl(\Gamma^\mu\bigr)_{~j}^{i}$, $i,j=0,1,2,3,4$ such that
(cf.~\cite[chapter~I, equations~(4.38) and~(4.37)]{Bogolyubov&Shirkov}):
	\begin{equation}	\label{8.2}
\varphi =
a
	\begin{pmatrix}
\ih\mathcal{D}_0\phi \\ \ih\mathcal{D}_1\phi \\
\ih\mathcal{D}_2\phi \\ \ih\mathcal{D}_3\phi \\ mc\phi
	\end{pmatrix},
\quad
\bigl(\Gamma^\mu\bigr)_{~j}^{i}
=	\begin{cases}
	1& 		\text{for $(i,j)=(\mu,4)$}	\\
	\eta_{\mu\mu}&	\text{for }(i,j)=(4,\mu)	\\
	0&		\text{otherwise}
	\end{cases}
	\end{equation}
where the complex constant $a\not=0$ is insignificant for us and can be
(partially) fixed by an appropriate normalization of $\varphi$.

	Then a simple checking shows that~\eref{8.1} is equivalent to
(cf.~\eref{7.1})
	\begin{equation}	\label{8.3}
(\ih\Gamma^\mu\mathcal{D}_\mu - mc\openone_5) \varphi = 0
	\end{equation}
with $\openone_5=\diag(1,1,1,1,1)$ being the $5\times5$ unit matrix, or
(cf.~\eref{7.22})
	\begin{equation}	\label{8.4}
\ih\Gamma^\mu\pd_\mu\varphi = \ope{K}\varphi,
\qquad
\ope{K} := mc\openone_5 + \frac{e}{c} \, \Gamma^\mu\ope{A}_\mu.
	\end{equation}

	Now it is evident that \emph{mutatis mutandis}, taking $\Gamma^\mu$
for $\gamma^\mu$, $\varphi$ for $\psi$, etc., the (bundle) machinery
developed in Sect.~\ref{Sect7} for the Dirac equation can be applied to the
Klein\ndash Gordon equation in the form~\eref{8.3}. Since the transferring of
the results obtained in Sect.~\ref{Sect7} for Dirac equation to Klein\ndash
Gordon one is absolutely trivial,%
\footnote{%
Both coincide up to notation or a meaning of the corresponding symbols.%
}
we are not going to present  here the bundle description of the latter
equation.

	Since the relativistic wave equations for particles with spin greater
than $1/2$ are versions or combinations of Dirac and Klein\ndash Gordon
equations~\cite{Nelipa,Bjorken&Drell-1,Messiah-2,Bogolyubov&Shirkov}, for them
is \emph{mutatis mutandis} applicable the bundle approach developed in
Sect.~\ref{Sect7} for Dirac equation or/and its version for
Klein\ndash Gordon  one pointed above.

\section
[Propagators and evolution transports or operators]
{Propagators and \\ evolution transports or operators}
\label{Sect9}

	The propagators, called also propagator functions or Green functions,
are solutions of the wave equations with point\nobreakdash-like unit source
and satisfy appropriate (homogeneous) boundary conditions corresponding to
a concrete problem under exploration~\cite{Bjorken&Drell-1,Itzykson&Zuber}.
Undoubtedly these functions play an important r\^ole in the mathematical
apparatus of (relativistic) quantum mechanics and its physical
interpretation~\cite{Bjorken&Drell-1,Itzykson&Zuber}. By this reason it is
essential to be investigated the connection between propagators and evolution
operators or/and transports. As we shall see below, the latter can be
represented as integral operators whose kernel is connected in a simple way
with the corresponding propagator.

\subsection{Green functions (review)}
\label{Subsect9.0}

	Generally~\cite[article ``Green function'']{Physicedia-1} the
\emph{Green function} $g(x',x)$ of a linear differential  operator $L$ (or of
the equation $Lu(x)=f(x)$) is the kernel of the integral operator inverse to
$L$. As the kernel of the unit operator is the Dirac delta\ndash function
$\delta^4(x'-x)$, the Green function is a fundamental solution of the
non\ndash homogeneous equation
	\begin{equation}	\label{9.a}
Lu(x)=f(x) ,
	\end{equation}
\ie treated as a generalized function $g(x',x)$ is a solution of
	\begin{equation}	\label{9.b}
L_{x'}g(x',x) = \delta^4(x'-x) .
	\end{equation}
Given a Geen function $g(x',x)$, the solution of~\eref{9.a} is
	\begin{equation}	\label{9.c}
u(x') = \int g(x',x)f(x) \,\od^4 x.
	\end{equation}

	A concrete Green function $g(x',x)$ for $L$ (or~\eref{9.a})
satisfies, besides~\eref{9.b}, certain (homogeneous) boundary conditions on
$x'$ with fixed $x$, \ie it is the solution of a fixed boundary\ndash value
problem for equation~\eref{9.b}. Hence, if $g_f(x',x)$ is some fundamental
solution, then
	\begin{equation}	\label{9.d}
g(x',x ) = g_f(x',x) + g_0(x',x)
	\end{equation}
where $g_0(x',x )$ is a solution of the homogeneous equation
$L_{x'}g_0(x',x ) = 0 $ chosen such that  $g(x',x )$ satisfies the required
boundary conditions.

	Suppose $g(x',x )$ is a Green function of $L$ for some boundary\ndash
value (or initial\ndash value) problem. Then, using~\eref{9.b}, we can verify
that
	\begin{equation}	\label{9.e}
L_{x'}\Bigl(\int g(x',x )u(x) \,\od^3\Vect{x}\Bigr)
= \delta(ct'-ct)u(x')
	\end{equation}
where $x=(ct,\Vect{x})$ and $x'=(ct',\Vect{x}')$. Therefore the
solution of the problem
	\begin{equation}	\label{9.f}
L_xu(x) = 0 , \quad u(ct_0,\Vect{x}) = u_0(\Vect{x})
	\end{equation}
is
	\begin{equation}	\label{9.g}
u(x) = \int g\bigl(x,(ct_0,\Vect{x}_0)\bigr)  u_0(\Vect{x}_0)
	\,\od^3 \Vect{x}_0,
\quad
x_0 = (ct_0,\Vect{x}_0)
\qquad\text{ for $t\not=t_0$} .
	\end{equation}
	From here we can make the conclusion that, if~\eref{9.f} admits an
evolution operator $\ope{U}$ such that (cf.~\eref{2.2})
	\begin{equation}	\label{9.h}
u(x) \equiv u(ct,\Vect{x})
=\ope{U}(t,t_0)u(ct_0,\Vect{x}) ,
	\end{equation}
then the r.h.s\ of~\eref{9.g} realizes $\ope{U}$ as an integral operator with
a kernel equal to the Green function $g$.

	Since all (relativistic or not) wave equations are versions
of~\eref{9.f}, the corresponding evolution operators, if any, and Green
functions (propagators) are connected as just described. Moreover, if some
wave equation does not admit (directly) evolution operator, e.g.\ if it is
of order greater than one, then we can \emph{define} it as the corresponding
version of the integral operator in the r.h.s.\ of~\eref{9.g}. In this way
is established a one\nobreakdash-to\nobreakdash-one onto correspondence
between the evolution operators and Green functions for any particular
problem like~\eref{9.f}.

	And a last general remark. The so-called S-matrix finds a lot of
applications in quantum theory
~\cite{Bjorken&Drell-1,Itzykson&Zuber,Bogolyubov&Shirkov}. By definition $S$
is an operator transforming the system's state vector
$\psi(-\infty,\Vect{x})$ before scattering (reaction) into the one
$\psi(+\infty,\Vect{x})$ after it:
\[
\lim_{t\to+\infty}\psi(ct,\Vect{x})
=: S \lim_{t\to-\infty}\psi(ct,\Vect{x}) .
\]
So, e.g.\ when~\eref{9.h} takes place, we have
	\begin{equation}	\label{9.j}
S
=  \lim_{t_\pm \to \pm\infty} \ope{U}(t_+,t_-)
=: \ope{U}(+\infty,-\infty) .
	\end{equation}

	Thus the above-mentioned connection between evolution operators and
Green functions can be used for expressing the S\ndash matrix in terms of
propagators. Such kind of formulae are often used in relativistic quantum
mechanics~\cite{Bjorken&Drell-1}.

\subsection {Nonrelativistic case (Schr\"odinger equation)}
\label{Subsect9.1}

	The (retarded) Green function $g(x',x)$, $x',x\in M$, for the
Schr\"odinger equation~\eref{2.1} is defined as the solution of the
boundary\ndash value problem~\cite[\S~22]{Bjorken&Drell-1}
	\begin{align}	\label{9.1}
\bigl[ \ih\frac{\pd}{\pd t'} - \Ham(x') \bigr] g(x',x) &= \delta^4(x'-x),
\\
\label{9.2}
g(x',x) &= 0 \qquad \text{for $t'<t$}
	\end{align}
where $x'=(ct',\Vect{x}')$ $x=(ct,\Vect{x})$,
 $\Ham(x)$ is system's Hamiltonian, and $\delta^4(x'-x)$ is the
4-dimensional (Dirac) $\delta$-function.

	Given $g$, the solution  $\psi(x')$ (for $t'>t$) of~\eref{2.1} is%
\footnote{%
One should not confuse the notation $\psi(x)=\psi(ct,\Vect{x})$,
$x=(ct,\Vect{x})$ of this section and $\psi(t)$ from Sect.~\ref{Sect2}.
The latter is the wavefunction at a moment $t$ and the former is its value at
the spacetime point $x=(ct,\Vect{x})$. Analogously,
\(
\Psi_\gamma(x)\equiv\Psi_\gamma(ct,\Vect{x})
:=l_{\gamma(t)}^{-1}\bigl(\psi(ct,\Vect{x})\bigr)
\)
 should not be confused with $\Psi_\gamma(t)$ from Sect.~\ref{Sect2}. A
notation like $\psi(t)$ and $\Psi_\gamma(t)$ will be used if the spatial
parts of the arguments are inessential, as in Sect.~\ref{Sect2}, and there is
no risk of ambiguities.%
}
	\begin{equation}	\label{9.3}
\theta(t'-t)\psi(x')
= \ih\int\ordinary^3\Vect{x} g(x',x)\psi(x)
	\end{equation}
where the $\theta$-function $\theta(s),\ s\in\mathbb{R}$, is defined by
$\theta(s)=1$ for $s>0$ and $\theta(s)=0$ for $s<0$.

	Combining~\eref{2.2} and~\eref{9.3}, we find the basic connection
between the evolution operator $\ope{U}$ and Green function of Schr\"odinger
equation:
	\begin{equation}	\label{9.4}
\theta(t'-t)\bigl[\ope{U}(t',t)\bigl(\psi(ct,\Vect{x}')\bigr)\bigr]
= \ih\int\ordinary^3\Vect{x}
g\bigl( (ct',\Vect{x}'),(ct,\Vect{x}) \bigr)
\psi(ct,\Vect{x}) .
	\end{equation}
Actually this formula, if $g$ is known, determines $\ope{U}(t',t)$ for all
$t'$ and $t$, not only for $t'>t$, as $\ope{U}(t',t)=\ope{U}^{-1}(t,t')$ and
$\ope{U}(t,t)=\id_\fibre$ with $\fibre$ being the system's Hilbert space
(see~\eref{2.2} or~\cite[sect.~2]{bp-BQM-introduction+transport}).
Consequently \emph{the evolution operator for the Schr\"odinger equation
can be represented as an integral operator whose kernel, up to the
constant $\ih$, is exactly the (retarded) Green function for it}.

	To write the bundle version of~\eref{9.4}, we introduce the
\emph{Green operator} which is simply a multiplication with the Green
function:
	\begin{equation}	\label{9.5}
\ope{G}(x',x) := g(x',x)\id_\fibre\colon\fibre\to\fibre.
	\end{equation}
The corresponding to it \emph{Green morphism} $G$ is given via
(see~\eref{2.11} and cf.~\eref{2.4})
	\begin{equation}	\label{9.6}
\mor{G}_\gamma(x',x)
:= l_{\gamma(t')}^{-1} \circ\ope{G}(x',x)\circ l_{\gamma(t)}
=  g(x',x)l_{\gamma(t')}^{-1}\circ l_{\gamma(t)}
\colon\fibreover{\gamma(t)}\to\fibreover{\gamma(t')} .
	\end{equation}

	Now, acting on~\eref{9.4} from the left by $l_{\gamma(t')}^{-1}$ and
using~\eref{2.3} and~\eref{2.4}, we obtain
	\begin{equation}	\label{9.7}
\theta(t'-t)
\bigl[\mor{U}_\gamma(t',t)\bigl(\Psi_\gamma(ct,\Vect{x}')\bigr)\bigr]
= \ih\int\ordinary^3\Vect{x}
\mor{G}_\gamma\bigl( (ct',\Vect{x}'),(ct,\Vect{x}) \bigr)
\Psi_\gamma(ct,\Vect{x}) .
	\end{equation}
Therefore for the Schr\"odinger equation the
\emph{%
	evolution transport $\mor{U}$ can be represented as an integral
	operator with kernel equal to $\ih$ times the Green morphism $G$%
}.

	Taking as a starting point~\eref{9.4} and~\eref{9.7}, we can obtain
different representations for the evolution operator and transport by applying
concrete formulae for the Green function. For example, if a complete set
$\{\psi_a(x)\}$ of orthonormal solutions of Schr\"odinger equation satisfying
the completeness condition%
\footnote{%
Here and below the symbol $\sum_{a}$ denotes a sum and/or integral over the
discrete and/or continuous spectrum. The asterisk ($*$) means complex
conjugation.%
}
\[
  \sum_a \psi_a(ct,\Vect{x}')\psi_a^*(ct,\Vect{x})
=\delta^3(\Vect{x}'- \Vect{x})
\]
is know, then~\cite[\S~22]{Bjorken&Drell-1}
\[
g(x',x)
= \iih\theta(t'-t) \sum_a \psi_a(x')\psi_a^*(x)
\]
which, when substituted into~\eref{9.4}, implies
	\begin{equation}	\label{9.8}
\ope{U}(t',t)\psi(ct,\Vect{x}')
=\sum_a \psi_a(x')
\int\od^3\Vect\,{x}\psi_a^*(ct,\Vect{x}) \psi(ct,\Vect{x}).
	\end{equation}
Note, the integral in this equation is equal to the  $a$-th coefficient
of the expansion of $\psi$ over $\{\psi_a\}$.

\subsection {Dirac equation}
\label{Subsect9.2}

	Since the Dirac equation~\eref{7.1} is a first-order linear partial
differential equation, it admits both evolution operator and Green
function(s) (propagator(s)). From a generic view\nobreakdash-point, the
only difference from the Schr\"odinger equation is that~\eref{7.1} is a
\emph{matrix} equation; so the corresponding Green functions are actually
Green matrices, \ie  Green matrix\ndash valued functions. Otherwise the
results of Subsect.~\ref{Subsect9.1} are \emph{mutatis mutandis} applicable
to the theory of Dirac equation.

	The (retarded) Green matrix (function) or propagator for Dirac
equation~\eref{7.1} is a $4\times4$ matrix\ndash valued function $g(x',x)$
depending on two arguments $x',x\in M$ and such
that~\cite[sect.~2.5.1 and~2.5.2]{Itzykson&Zuber}
	\begin{align}
\label{9.20}
(\ih \Slashed{D}_{x'} - mc\openone_4) g(x',x) &= \delta(x'-x)
\\
\label{9.21}
g(x',x) &= 0 \qquad \text{for $t<t'$}.
        \end{align}

	For a free Dirac particle, \ie for $\Slashed{D}=\slashed{\pd}$ or
$\ope{A}_\mu=0$, the explicit expression $g_0(x',x)$ for $g(x',x)$ is
derived in~\cite[sect.~2.5.1]{Itzykson&Zuber}, where the notation
$\mathcal{K}$ instead of $g_0$ is used. In an external electromagnetic field
$\ope{A}_\mu$ the Green matrix $g$ is a solution of the integral equation%
\footnote{%
The derivation of~\eref{9.22} is the same as for the Feynman propagators
$S_F$ and $S_A$ given in~\cite[sect.~2.5.2]{Itzykson&Zuber}. The propagators
$S_F$ and $S_A$ correspond to $g_0$ and $g$ respectively, but satisfy other
boundary conditions~\cite{Itzykson&Zuber,Bjorken&Drell-1}.%
}
	\begin{equation}	\label{9.22}
g(x',x)
= g_0(x',x) + \int\od^4y\, g_0(x',y) \frac{e}{c}\Slashed{\ope{A}}(y) g(y,x)
	\end{equation}
which includes the corresponding boundary condition.%
\footnote{%
Due to~\eref{9.d}, the integral equation~\eref{9.22} is valid for any Green
function (matrix) of the Dirac equation~\eref{7.1}.%
}
The iteration of this equation results in the perturbation series for $g$
(cf.~\cite[sect.~2.5.2]{Itzykson&Zuber}).

	If the (retarded) Green matrix $g(x',x)$ is known, the solution
$\psi(x')$ of Dirac equation (for $t'>t$) is
	\begin{equation}	\label{9.23}
\theta(t'-t)\psi(x')
= \ih\int\od^3\Vect{x}\, g(x',x)\gamma^0\psi(x) .
	\end{equation}

	Hence, denoting by $\ope{U}$ the non\ndash relativistic (see
Sect.~\ref{Sect4}) Dirac evolution operator, from the equations~\eref{2.2},
and~\eref{9.23}, we find:
	\begin{equation}	\label{9.24}
\theta(t'-t)
\bigl[ \ope{U}(t',t)\bigl(\psi(ct,\Vect{x}') \bigr)\bigr]
= \ih\int\od^3\Vect{x}\,
	g\bigl((ct',\Vect{x}'),(ct,\Vect{x})\bigr)
	\gamma^0\psi(ct,\Vect{x}) .
	\end{equation}
So, the evolution operator admits an integral representation
whose kernel, up to the right multiplication with  $\ih\gamma^0$, is equal to
the (retarded) Green function for Dirac equation.

	Similarly to~\eref{9.7}, now the bundle version of~\eref{9.24} is
	\begin{align}
		\label{9.25a}
\theta(t'-t)
\bigl[ \mor{U}(t',t)
	\bigl(\Psi_\gamma(ct,\Vect{x}')\bigr) \bigr]
& = \ih\int\od^3\Vect{x}\,
	\mor{G}_\gamma(x',x) G^0(\gamma(t))\Psi_\gamma(ct,\Vect{x}),
	\end{align}
where $G^0(x)$ is defined by~\eref{7.17} with $\mu=0$ and (cf.~\eref{9.6})
	\begin{equation}	\label{9.26}
\mor{G}_\gamma(x',x)
:= l_{\gamma(t')}^{-1}\circ \ope{G}(x',x) \circ l_{\gamma(t)},
\quad
\mor{G}(x',x)
:= l_{x'}^{-1}\circ \ope{G}(x',x) \circ l_{x}
	\end{equation}
with (\cf.~\eref{9.5})
	\begin{equation}	\label{9.27}
\ope{G}(x',x) = g(x',x) \id_\fibre .
	\end{equation}
(Here $\fibre$ is the space of 4-spinors.)

	Analogously to the above results, one can obtained such for other
propagators, \eg for the Feynman one~\cite{Bjorken&Drell-1,Itzykson&Zuber},
but we are not going to do this here as it is a trivial variant of the
procedure described.

\subsection {Klein-Gordon equation}
\label{Subsect9.3}

	The (retarded) Green function $g(x',x)$ for the Klein-Gordon
equation~\eref{8.1} is a solution to the boundary\ndash value
problem~\cite[sect.~1.3.1]{Itzykson&Zuber}
	\begin{align}
\label{9.31}
\Bigl.\Bigl(
\mathcal{D}^\mu\mathcal{D}_\mu + \frac{m^2c^2}{\hbar^2}
	\Bigr)\Bigr|_{x'} g(x',x)
&= \delta^4(x'-x),
\\
\label{9.32}
g(x',x) &= 0 \qquad \text{for $t'<t$}.
	\end{align}
For a free particle its explicit form can be found
in~\cite[sect.~1.3.1]{Itzykson&Zuber}.

	A simple verification proves that, if $g(x',x)$ is known, the
solution $\phi$ of~\eref{8.1} (for $t'>t$) is given by ($x^0=ct$)
	\begin{equation}	\label{9.33}
\theta(t'-t)\phi(x')
= \mspace{-1.23mu}
  \int\od^3\Vect{x}\,
	\Bigl[ \frac{\pd g(x',x)}{\pd x^0} \phi(x)
	+ g(x',x) \Bigl( 2\frac{\pd\phi(x)}{\pd x^0}
		-\frac{e}{\ih c}\ope{A}^0(x)\phi(x) \Bigr)
	\Bigr] .
	\end{equation}

	Introducing the matrices
	\begin{equation}	\label{9.34}
\psi(x)
:= 	\begin{pmatrix} \phi(x) \\ \pd_0|_x\phi(x) \end{pmatrix},
\qquad
\mathsf{g}(x',x)
:=	\begin{pmatrix} \mathcal{D}_0|_x g(x',x) \\ 2g(x',x) \end{pmatrix},
	\end{equation}
we can rewrite~\eref{9.33} as
	\begin{equation}	\label{9.35}
\theta(t'-t)\phi(x')
= \int\od^3\Vect{x}\,
	\mathsf{g}^\top(x',x)\cdot\psi(x)
	\end{equation}
where the dot ($\cdot$) denotes matrix multiplication.

	An important observation is that for $\psi$  the Klein-Gordon equation
transforms into first\ndash order Schr\"odinger\ndash type equation (see
Sect.~\ref{Sect5}) with Ha\-miltonian
$\lindexrm[\Ham]{\mspace{32mu}c}{K-G}$ given by~\eref{5.3} in which
$\id_{\ldots}$ is replaced by $c\id_{\ldots}$.

	Denoting the (retarded) Green function, which is in fact $2\times2$
matrix, and the evolution operator for this equation by
\(
\widetilde{\ope{G}}(x',x) =
\bigl[ \widetilde{\ope{G}}_{~b}^{a}(x',x)\bigr]_{a,b=1}^{2}
\)
and
\(
\widetilde{\ope{U}}(t',t) =
\bigl[ \widetilde{\ope{U}}_{~b}^{a}(t',t)\bigr]_{a,b=1}^{2},
\)
respectively, we see that (\cf Subsect.~\ref{Subsect9.1},
equation~\eref{9.3})
	\begin{equation}	\label{9.36}
\theta(t'-t) \bigl[\widetilde{\ope{U}}(t',t) \psi(ct,\Vect{x})\bigr]
= \theta(t'-t)\psi(x')
= \int\od^3\Vect{x}\,
	\widetilde{\ope{G}}(x',x) \psi(x) .
	\end{equation}
Comparing this equations with~\eref{9.34} and~\eref{9.35}, we find
\[
\bigl(
\widetilde{\ope{G}}_{~1}^{1}(x',x),\widetilde{\ope{G}}_{~2}^{1}(x',x) \bigr)
= \mathsf{g}^\top(x',x) .
\]
The other matrix elements of $\widetilde{\ope{G}}$ can also be connected with
$\mathsf{g}(x',x)$ and its derivatives, but this is inessential for the
following.

	In this way we have connected, via~\eref{9.36}, the evolution
operator and the (retarded) Green function for a concrete first\ndash order
realization of Klein\ndash Gordon equation. It is almost evident that this
procedure \emph{mutatis mutandis} works for any such realization; in every
case the corresponding Green function (resp.\ matrix) being a (resp.\
matrix\ndash valued) function of the Green function $g(x',x)$ introduce
via~\eref{9.31} and~\eref{9.32}. For instance, the treatment of the 5\ndash
dimensional realization given by~\eref{8.2} and~\eref{8.3} is practically
identical to the one of Dirac equation in Subsect.~\ref{Subsect9.2}, only the
$\gamma$\ndash matrices $\gamma^\mu$ have to be replace with the $5\times5$
matrices $\Gamma^\mu$ (defined by~\eref{8.2}). This results in a $5\times5$
matrix evolution operator $\ope{U}(x',x)$, etc.

	Since the bundle version of~\eref{9.36} or an analogous result for
$\ope{U}(x',x)$ is absolutely trivial (\cf Subsect.~\ref{Subsect9.1}
and~\ref{Subsect9.2} resp.), we are not going to write it here; up to the
meaning of notation it coincides with~\eref{9.7} or~\eref{9.25a}
respectively.

\section {Conclusion}
\label{II.Conclusion}

	In this investigation we have reformulated the relativistic wave
equations in terms of fibre bundles. In the bundle formulation the
wavefunctions are represented as (state) liftings of paths or sections along
paths (time\ndash dependent approach) or simply sections (covariant approach)
of a suitable vector bundle over the spacetime. The covariant approach,
developed in the present work, has an advantage of being explicitly covariant
while in the time\ndash dependent one the time plays a privilege r\^ole. In
both cases the evolution (in time or in spacetime resp.) is described via
a linear transport in the bundle mentioned. The state
liftings or sections are linearly transported by means of the corresponding
(evolution) transports. We have also explored some links between evolution
operators or transports and the retarded Green functions (or matrices) for
the corresponding wave equations: the former turn to have realization as
integral operators whose kernel is equal to the latter ones up to a
multiplication with a constant complex number or matrix.

	These connections suggest the idea for introducing `retarded', or, in
a sense, `causal' evolution operators or transports as a product of the
evolution operators or transports with $\theta$\ndash function of the
difference of the times corresponding to the first and second arguments of the
transport or operator.

	Most of the possible generalizations of the bundle non\ndash
relativistic quantum mechanics, pointed
in~\cite{bp-BQM-interpretation+discussion}, are \emph{mutatis mutandis} valid
with respect to the bundle version of relativistic quantum mechanics,
developed in the present work. The only essential change is that, in the
relativistic region, the spacetime model is fixed as the Minkowski spacetime.

	A further development of the ideas presented in this investigation
leads to their application to (quantum) field theory which will be done
elsewhere.


\appendix
\renewcommand{\thesection}{Appendix~\Alph{section}}

\section
[\hspace*{4.75em}Linear transports along maps in fibre bundles]
{Linear transports along maps in\\ fibre bundles}

	\renewcommand{\thesection}{\Alph{section}}
	\label{AppendixA}

	In this appendix we recall a few simple facts concerning (linear)
transports along maps, in particular along paths, required for the present
investigation. The below\ndash presented material is abstracted
from~\cite{bp-TM-general,bp-normalF-LTP,bp-LTP-general} where further details can be found
(see also~\cite[sect.~3]{bp-BQM-introduction+transport}).

	Let $(E,\pi,B)$ be a topological bundle with base $B$,
bundle (total, fibre) space $E$, projection $\pi:E\to B$, and homeomorphic
fibres $\pi^{-1}(x),\ x\in B$.  Let the set $N$ be not empty,
$N\neq\varnothing$, and there  be given a map $\varkappa:N\to B$. By $\id_X$
is denoted the identity map of a set $X$.

	\begin{Defn}	\label{DefnLT.1}
	A transport along maps in the bundle $(E,\pi,B)$ is a map
$K$ assigning to any map $\varkappa:N\to B$ a map  $K^\varkappa$,
transport along $\varkappa$, such that
$K^\varkappa:(l,m)\mapsto K_{l\to m}^{\varkappa}$, where for
every $l,m\in N$ the map
\begin{equation}	\label{LT.1}
K_{l\to m}^{\varkappa}:\pi^{-1}(\varkappa(l))\to \pi^{-1}(\varkappa(m)),
\end{equation}
called transport along  $\varkappa$ from  $l$ to $m$, satisfies the
equalities:
\begin{align}
K_{m\to n}^{\varkappa} \circ  K_{l\to m}^{\varkappa} &=
	K_{l\to n}^{\varkappa} , & l,m,n\in N, 		\label{LT.2}\\
K_{l\to l}^{\varkappa} &= \id_{\pi^{-1}(\varkappa(l))},
				 & l\in N.		\label{LT.3}
\end{align}
\end{Defn}

	If $(E,\pi,B)$ is a complex (or real) vector bundle and the
maps~\eref{LT.1} are linear, \ie
	\begin{equation}      \label{LT.4}  \!
K_{l\to m}^{\varkappa}(\lambda u + \mu v) =
\lambda K_{l\to m}^{\varkappa}u + \mu K_{l\to m}^{\varkappa}v,\ \>
\lambda,\mu\in\mathbb{C\mathrm{\ (or\ }R)},\ \>
u,v\in\pi^{-1}(\varkappa(l)),
	\end{equation}
the transport $K$ is called \emph{linear}. If $\varkappa$ belongs to the set
of paths in $B$,
\(
\varkappa\in\{\gamma\colon J\to B,\
	J\text{ being }\mathbb{R}\text{-interval}\},
\)
we said that the transport $K$ is \emph{along paths}.

	For the present work is important that the class of linear transports
along the identity map $\id_B$ of $B$ coincides with the class of \emph{flat}
linear transports along paths%
\footnote{%
The \emph{flat} linear transports along paths are defined as ones with
vanishing curvature operator~\cite[sect.~2]{bp-LTP-Cur+Tor-prop}.
By~\cite[theorem.~6.1]{bp-LTP-Cur+Tor-prop} we can equivalently define them by
the property that they depend only on their initial and final points, \ie if
$K_{s\to t}^{\gamma}$, $\gamma\colon J\to B$, $s,t\in J\subset\mathbb{R}$,
depends only on $\gamma(s)$ and $\gamma(t)$ but not on the path $\gamma$
itself.%
}
(see the comments after equation~(2.3) of~\cite{bp-TM-general}).

	The general form of a transport along maps is described by the
following result.
	\begin{Thm}	\label{ThmLT.1}
Let $\varkappa:N\to B$. The map
$K:\varkappa\mapsto K^\varkappa:(l,m)\mapsto K_{l\to m}^{\varkappa}$,
$l,m\in N$
is a transport along $\varkappa$ if and only if there exist a set
$Q$ and a family of bijective maps
 $\{ F_{n}^{\varkappa}:\pi^{-1}(\varkappa(n))\to Q,\ n\in N \}$
such that
	\begin{equation}	\label{LT.5}
K_{l\to m}^{\varkappa} = \left(F_{m}^{\varkappa}\right)^{-1} \circ
\left(F_{l}^{\varkappa}\right),	\quad l,m\in N.
	\end{equation}
The maps $F_{n}^{\varkappa}$ are defined up to a left composition with
bijective map depending only on $\varkappa$, i.e.~\eref{LT.5} holds for given
families of maps
$\{ F_{n}^{\varkappa}:\pi^{-1}(\varkappa(n))\to Q,\ n\in N \}$ and
\(
\{ ^\prime\! F_{n}^{\varkappa}:\pi^{-1}(\varkappa(n))\to
{}^\prime\! Q,\ n\in N \}
\)
for some sets $Q$ and  $^\prime\! Q$ iff there is a bijective map
$D^\varkappa:Q\to{}^\prime\! Q$ such that
	\begin{equation}	\label{LT.6}
^\prime\! F_{n}^{\varkappa} = D^\varkappa \circ F_n^\varkappa,\quad n\in N.
	\end{equation}
	\end{Thm}

	For the purposes of this investigation we need a slight
generalization of~\cite[definition~4.1]{bp-TM-general}, viz. we want to
replace in it $\mathbf{N}\subset\mathbb{R}^k$ with an arbitrary
differentiable manifold. Let $N$ be a differentiable manifold and
$\{x^a\ :\ a=1,\dots,\dim{N}\}$ be coordinate system in a neighborhood of
$l\in N$. For
$\varepsilon\in(-\delta,\delta)\subset\mathbb{R}$, $\delta\in\mathbb{R}_+$
and  $l\in N$ with coordinates $l^a=x^a(l)$, we define
$l_b(\varepsilon)\in N$, $b=1,\dots,\dim{N}$ by
\(
l_{b}^{a}(\varepsilon):=x^a(l_b(\varepsilon)):=l^a +
	\varepsilon \delta_{b}^{a}
\)
where the Kroneker $\delta$\ndash symbol is given by $\delta_{b}^{a}=0$ for
$a\not=b$ and $\delta_{b}^{a}=1$ for $a=b$.
	Let $\xi=(E,\pi,B)$ be a vector bundle,
$\varkappa\colon N\to B$ be injective (\ie 1:1 mapping),
and $\mathrm{Sec}^p(\xi)$ (resp.\ $\mathrm{Sec}(\xi)$) be the
set of $C^p$ (resp.\ all) sections over $\xi$.
Let $L_{l\to m}^{\varkappa}$ be a $C^1$ (on $l$) \emph{linear} transport along
$\varkappa$.
Now the modified definition reads:%
\footnote{%
We present below, in definition~\ref{DefnLT.2}, directly the definition of a
\emph{section}\ndash derivation along injective mapping $\varkappa$ as only
it will be employed in the present paper (for $\varkappa=\id_N$). If
$\varkappa$ is not injective, the mapping~\eref{LT.7} could be
multiple\ndash valued at the points of self\ndash intersection of
$\varkappa$, if any. For some details when $\varkappa$ is an arbitrary path
in $N$, see~\cite[subsect.~3.3]{bp-BQM-introduction+transport}.%
}
	\begin{Defn}	\label{DefnLT.2}
	The $a$-th, $1\le a\le\dim N$, partial
(section\ndash)derivation along maps generated by  $L$ is a  map
${_a  }\mathcal{D}:\varkappa\mapsto{_a  }\mathcal{D}^\varkappa$
where the $a$-th (partial) derivation ${_a  }\mathcal{D}^\varkappa$
along $\varkappa$ (generated by $L$)  is  a map
	\begin{equation}	\label{LT.7}
{_a  }\mathcal{D}^\varkappa : l \mapsto \mathcal{D}^\varkappa_{l^a} ,
	\end{equation}
where the $a$-th (partial) derivative $\mathcal{D}^\varkappa_{l^a}$ along
$\varkappa$ at $l$ is a map
	\begin{equation}	\label{LT.9}
	\mathcal{D}_{l^a}^{\varkappa} :
\mathrm{Sec}^1\left(\left.\xi\right|_{\varkappa(N)}\right) \to
\pi^{-1}(\varkappa(l))
	\end{equation}
defined for
$\sigma\in\mathrm{Sec}^1\left(\xi|_{\varkappa(N)}\right)$ by
	\begin{equation} 	\label{LT.8}
\left({_a  }\mathcal{D}^\varkappa\sigma\right) (\varkappa(l)) :=
\lim_{\varepsilon\to 0}  	\left[ \frac{1}{\varepsilon} \bigl(
L_{l_a(\varepsilon) \to l}^{\varkappa}\sigma(\varkappa(l_a(\varepsilon)))
				- \sigma(\varkappa(l)) \bigr) \right].
	\end{equation}
	\end{Defn}

	Accordingly can be modified the other definitions
of~\cite[sect.~4]{bp-TM-general}, all the results of it being
\emph{mutatis mutandis} valid. In particular, we have:
	\begin{Prop}	\label{PropLT.1}
	The operators $\mathcal{D}^\varkappa_{l^a}$ are
($\mathbb{C}$-)linear
and
	\begin{equation}	\label{LT.10}
\mathcal{D}_{m^a}^{\varkappa} \circ L_{l\to m}^{\varkappa} \equiv 0 .
	\end{equation}
	\end{Prop}

	\begin{Prop}	\label{PropLT.2}
	If
$\sigma\in\mathrm{Sec}^1\left(\xi|_{\varkappa(N)}\right)$,
then
	\begin{equation}	\label{LT.11}
	\mathcal{D}_{l^a}^{\varkappa} \sigma
= \sum_i
\Bigl[
	\frac{\partial\sigma^i(\varkappa(l))}{\partial l^a}
	+ \sum_{j} {_a  }\Gamma_{\;j}^{i}(l;\varkappa)\sigma^j(\varkappa(l))
\Bigr] e_i(l) ,
	\end{equation}
where $\{e_i(l)\}$ is a basis in $\pi^{-1}(\varkappa(l))$,
 $\sigma(\varkappa(l))=:\sum_{i}\sigma^i(\varkappa(l)) e_i(l)$,
and the \textbf{coefficients} of  $L$ are defined by
	\begin{equation}	\label{LT.12}
{_a  }\Gamma_{\;j}^{i}(l;\varkappa) :=
\left.
\frac{\partial L_{\;j}^{i}(l,m;\varkappa)}{\partial m^a}
\right|_{m=l} =
- \left.
\frac{\partial L_{\;j}^{i}(m,l;\varkappa)}{\partial m^a}
\right|_{m=l}.
	\end{equation}
Here $L_{\;i}^{j}(\cdots)$ are the components of $L$,
$L_{l\to m}^{\varkappa}e_i(l)=:\sum_j L_{\;i}^{j}(m,l;\varkappa)e_j(m)$.
	\end{Prop}

	The above general definitions and results will be used in this work
in the special case of linear transports along the identity map of the
bundle's base.

\renewcommand{\thesection}{Appendix~\Alph{section}}

\section[\hspace*{4.75em}Matrix operators]{Matrix operators}

	\renewcommand{\thesection}{\Alph{section}}
	\label{AppendixB}

	In this appendix we point to some peculiarities of linear
(matrix) operators acting on $n\times1$, $n\in\mathbb{N}$, matrix fields
over the space\nobreakdash-time $\basesp$. Such operators appear naturally in
the theory of Dirac equation where one often meets $4\times4$ matrices whose
elements are operators; e.g.\ an operator of this kind is
\(
\slashed{\partial} := \gamma^\mu\partial_\mu =
\bigl[ (\gamma^\mu)_{~\beta}^{\alpha}\partial_\mu \bigr]_{\alpha,\beta=0}^{3}
\)
where $\gamma^\mu$ are the well known Dirac $\gamma$\ndash
matrices~\cite{Itzykson&Zuber,Bjorken&Drell-1}.

	We call an $n\times n$, $n\in\mathbb{N}$ matrix
$B=[b_{~\beta}^{\alpha}]_{\alpha,\beta=1}^{n}$
a \emph{(linear) matrix operator}%
\footnote{%
Also a good term for such an object is (linear) \emph{matrixor}.%
}
if $b_{~\beta}^{\alpha}$ are (linear) operators acting on the space
$\mathit{K}^1$ of $C^1$ functions $f\colon M\to\mathbb{C}$.
If $\{ f_{\nu}^{0} \}$ is a basis in the set $M(n,1)$ of $n\times1$ matrices
with the $\mu$\nobreakdash-th element of $f_{\nu}^{0}$ being
$(f_{\nu}^{0})^\mu:=\delta_{\nu}^{\mu}$,%
\footnote{%
$\delta_{\nu}^{\mu}=0$ for $\mu\not=\nu$ and
$\delta_{\nu}^{\mu}=1$ for $\mu=\nu$. From here to equation~\eref{mo.9} in
this appendix the Greek indices run from 1 to $n\ge1$.%
}
then by definition
	\begin{equation}	\label{mo.1}
B\psi :=B(\psi) := B\cdot(\psi) :=
\sum_{\alpha,\beta=1}^{n}
\bigl( b_{~\beta}^{\alpha} (\psi_{0}^{\beta} \bigr)  f_{\alpha}^{0},
\qquad
\psi = \psi_{0}^{\beta} f_{\beta}^{0} \in M(n,1).
	\end{equation}
For instance, we have
\(
\slashed{\pd}\psi
= \sum_{\alpha,\beta,\mu=0}^{3}
  (\gamma^\mu)_{~\beta}^{\alpha}(\pd_\mu\psi_{0}^{\beta}) f_{\alpha}^{0}.
\)

	To any constant matrix $C=[c_{~\beta}^{\alpha}]$,
$c_{~\beta}^{\alpha}\in\mathbb{C}$, corresponds a matrix operator
 $\overline{C}:=[c_{~\beta}^{\alpha}\id_{\mathit{K}^1}]$. Since
$C\psi\equiv\overline{C}\psi$ for any $\psi$, we identify $C$ and
$\overline{C}$ and will make no difference between them.

	The multiplication of matrix operators, denoted by $\odot$, is a
combination of matrix multiplication (denoted by $\cdot$) and maps
(operators) composition (denoted by $\circ$). If $A=[a_{~\beta}^{\alpha}]$
and $B=[b_{~\beta}^{\alpha}]$ are matrix operators, the product of $A$
and $B$ is also a matrix operator such that
	\begin{equation}	\label{mo.2}
AB := A\odot B :=
\biggl[ \sum_{\mu}^{} a_{~\mu}^{\alpha} \circ b_{~\beta}^{\mu} \biggr] .
	\end{equation}
	One can easily show that this is an associative operation. It is
linear in the first argument and, if its first argument is linear matrix
operator, then it is linear in its second argument too. For constant matrices
(see above), the multiplication~\eref{mo.2} coincides with the usual matrix
one.

	Let $\{f_{\alpha}(x)\}$ be a basis in $M(n,1)$ depending on $x\in M$
and $f(x):=[f_{\alpha}^{\beta}(x)]$ be defined by the expansion
$f_{\alpha}(x)=f_{\alpha}^{\beta}(x)f_{\beta}^{0}$.
The \emph{matrix of a matrix operator} $B=[b_{~\beta}^{\alpha}]$ with respect
to $\{f_\alpha\}$, \ie the matrix of the matrix elements of $B$ considered as
an operator, is also a matrix operator
$\Mat{B}:=[B_{~\beta}^{\alpha}]$  such that
	\begin{equation}	\label{mo.3}
B\psi|_x
=: \sum_{\alpha,\beta}^{}
    \bigl( B_{~\beta}^{\alpha}(\psi^\beta) \bigr) \bigl.\bigr|_x f_\alpha(x),
\qquad
\psi(x) = \psi^\beta(x)f_\beta(x).
        \end{equation}
Therefore
	\begin{equation}	\label{mo.3'}
\varphi = B\psi
\iff
\Mat{\varphi} = \Mat{B}\Mat{\psi},
	\end{equation}
where $\Mat{\psi}\in M(n,1)$ is the matrix of the components of $\psi$
in the basis given. Comparing~\eref{mo.1} and~\eref{mo.3}, we get
	\begin{equation}	\label{mo.4}
B_{~\beta}^{\alpha}
= \sum_{\mu,\nu} \bigl( f^{-1}(x) \bigr)_{\mu}^{\alpha} \,
	b_{~\nu}^{\mu} \circ
	\bigl( f_{\beta}^{\nu}(x)\id_{\mathit{K}^1} \bigr) .
	\end{equation}

	For any matrix $C=[c_{~\beta}^{\alpha}]\in GL(n,\mathbb{C})$,
considered as a matrix operator (see above), we have
	\begin{equation}	\label{mo.5}
\Mat{C} = \bigl[C_{~\beta}^{\alpha}(x)\bigr],
\quad
C_{~\beta}^{\alpha}(x) =
\bigl( f^{-1}(x) \bigr)_{\mu}^{\alpha} \, c_{~\nu}^{\mu} f_{\beta}^{\nu},
\quad
C(f_\beta) = C_{~\beta}^{\alpha} f_\alpha .
	\end{equation}
Therefore, as one can expect, the matrix of the unit matrix
$\openone_n=\bigl[\delta_{\alpha}^{\beta}\bigr]_{\alpha,\beta=1}^{n}$
is exactly the unit matrix,
	\begin{equation}	\label{mo.6}
\pmb{\openone}_n = \openone_n .
	\end{equation}
If $\{f_\mu(x)\}$ does not depend on $x$, \ie if $f_{\alpha}^{\beta}$ are
complex constants, and $b_{~\nu}^{\mu}$ are linear, than~\eref{mo.4} implies
	\begin{equation}	\label{mo.7}
B_{~\beta}^{\alpha}
=
\bigl(f^{-1}(x)\bigr)_{\mu}^{\alpha} \, f_{\beta}^{\nu}(x) b_{~\nu}^{\mu}
\ \text{ for }\
\pd f_{\alpha}^{\beta}(x)/\pd x^\mu = 0 .
	\end{equation}
In particular, for a linear matrix operator $B$, we have
	\begin{equation}	\label{mo.8}
\Mat{B} = B \ \text{ for }\  f_{\beta}^{\alpha}
	= \delta_{\beta}^{\alpha}
	\ \text{ (\ie for $f_\alpha(x)=f_{\alpha}^{0}$)} .
	\end{equation}
Combining~\eref{mo.2} and~\eref{mo.4}, we deduce that the matrix of a product
of matrix operators is the product of the corresponding matrices:
	\begin{equation}	\label{mo.9}
C=A\odot B \quad\iff\quad
\Mat{C} = \Mat{A}\Mat{B} = \Mat{A}\odot\Mat{B}.
	\end{equation}

	After a simple calculation, we find the matrices of
$\openone_4\pd_\mu$ and $\slashed{\pd}=\gamma^\mu\pd_\mu$:%
\footnote{Here and below the Greek indices run from 0 to 3.}
	\begin{equation}	\label{mo.10}
\Mat{\pd}_\mu = \openone_4\pd_\mu + \ope{E}_\mu(x), \qquad
\Mat{\slashed{\pd}}
=
\ope{G}^\mu(x) \bigl( \openone_4\pd_\mu + \ope{E}_\mu(x) \bigr)
=
\ope{G}^\mu(x) \Mat{\pd}_\mu
	\end{equation}
where
\(
\ope{G}^\mu(x) = \bigl[ \ope{G}_{~\alpha}^{\lambda\mu}(x)
\bigr]_{\alpha,\lambda=0}^{3}
\)
and
\(
\ope{E}_\mu(x) = \bigl[ \ope{E}_{\mu\alpha}^{\lambda}(x)
\bigr]_{\alpha,\lambda=0}^{3}
\)
are defined via the expansions
	\begin{equation}	\label{mo.11}
\gamma^\mu f_\nu(x)	=: \ope{G}_{~\nu}^{\lambda\mu}(x) f_\lambda(x),
\quad
\pd_\mu f_\nu(x)	=: \ope{E}_{\mu\nu}^{\lambda}(x) f_\lambda(x) .
	\end{equation}
So, $\ope{G}^\mu$ is the matrix of $\gamma^\mu$ considered as a matrix
operator (see~\eref{mo.5}).%
\footnote{%
We denote the matrix of $\gamma^\mu$ by $\ope{G}^\mu$ instead of
$\boldsymbol{\gamma}^\mu(x)$ as usually~\cite{Bjorken&Drell-1,Itzykson&Zuber}
by $\boldsymbol{\gamma}$ is denoted the matrix 3\ndash vector
$(\gamma^1,\gamma^2,\gamma^3)$.%
}
Evidently
	\begin{equation}	\label{mo.12}
\Mat{\slashed{\pd}} = \slashed{\pd} \ \text{ for }\
	f_\alpha(x)=f_{\alpha}^{0} .
	\end{equation}

	Combining~\eref{mo.5},\eref{mo.6}, and~\eref{mo.10}, we get that the
matrix of the matrix operator $\ih\Slashed{D}-mc\openone_4$, entering in the
Dirac equation~\eref{7.1}, is
	\begin{equation}	\label{mo.13}
\ih\Mat{\Slashed{D}}-mc\pmb{\openone}_4 =
\ih\ope{G}^\mu(x)\bigl(\openone_4D_\mu + \ope{E}_\mu(x)\bigr) - mc\openone_4,
	\end{equation}
so that
	\begin{equation}	\label{mo.14}
\ih\Mat{\Slashed{D}}-mc\pmb{\openone}_4 =
\ih\Slashed{D}-mc\openone_4
\ \text{ for }\
f_\alpha(x) = f_{\alpha}^{0}.
	\end{equation}

\addcontentsline{toc}{section}{References}
\bibliography{bozhopub,bozhoref}

\begin{thebibliography}{10}

\bibitem{bp-BRQM-time-dependent}
Bozhidar~Z. Iliev.
\newblock Fibre bundle formulation of relativistic quantum mechanics. {I}.
  {Time}-dependent approach.
\newblock \\ LANL arXiv server, E-print No.\ quant-ph/0105056, 1998, May 2001.

\bibitem{bp-BQM-interpretation+discussion}
Bozhidar~Z. Iliev.
\newblock Fibre bundle formulation of nonrelativistic quantum mechanics. {V}.
  interpretation, summary, and discussion.
\newblock {\em International Journal of Modern Physics~A}, 17(2):245--258,
  2002.
\newblock \\ LANL arXiv server, E-print No.\ quant-ph/9902068, 1999.

\bibitem{bp-BQM-full}
Bozhidar~Z. Iliev.
\newblock Fibre bundle formulation of nonrelativistic quantum mechanics (full
  version).
\newblock \\ LANL arXiv server, E-print No.\ quant-ph/0004041, 2000, April
  2000.

\bibitem{Itzykson&Zuber}
C.~Itzykson and J.-B. Zuber.
\newblock {\em Quantum field theory}.
\newblock McGraw-Hill Book Company, New York, 1980.
\newblock Russian translation (in two volumes): Mir, Moscow, 1984.

\bibitem{Messiah-2}
A.~M.~L. Messiah.
\newblock {\em Quantum mechanics}, volume~II.
\newblock North Holland, Amsterdam, 1962.
\newblock Russian translation: Nauka, Moscow, 1979.

\bibitem{Bjorken&Drell-1}
J.~D. Bjorken and S.~D. Drell.
\newblock {\em Relativistic quantum mechanics}, volume~1.
\newblock McGraw-Hill Book Company, New York, 1964.
\newblock Russian translation: Nauka, Moscow, 1978.

\bibitem{bp-BQM-introduction+transport}
Bozhidar~Z. Iliev.
\newblock Fibre bundle formulation of nonrelativistic quantum mechanics. {I}.
  {Introduction}. {The} evolution transport.
\newblock {\em Journal of Physics A: Mathematical and General},
  34(23):4887--4918, 2001.
\newblock \\ LANL arXiv server, E-print No.\ quant-ph/9803084, 1998.

\bibitem{Messiah-1}
A.~M.~L. Messiah.
\newblock {\em Quantum mechanics}, volume~I.
\newblock North Holland, Amsterdam, 1961.
\newblock Russian translation: Nauka, Moscow, 1978.

\bibitem{bp-LTP-appl}
Bozhidar~Z. Iliev.
\newblock Linear transports along paths in vector bundles. {II}.~{Some}
  applications.
\newblock JINR Communication E5-93-260, Dubna, 1993.

\bibitem{bp-BQM-equations+observables}
Bozhidar~Z. Iliev.
\newblock Fibre bundle formulation of nonrelativistic quantum mechanics. {II}.
  {Equations} of motion and observables.
\newblock {\em Journal of Physics A: Mathematical and General},
  34(23):4919--4934, 2001.
\newblock \\ LANL arXiv server, E-print No.\ quant-ph/9804062, 1998.

\bibitem{Bogolyubov&Shirkov}
N.~N. Bogolyubov and D.~V. Shirkov.
\newblock {\em Introduction to the theory of quantized fields}.
\newblock Nauka, Moscow, third edition, 1976.
\newblock In Russian. English translation: Wiley, New York, 1980.

\bibitem{Nelipa}
N.~F. Nelipa.
\newblock {\em Physics of elementary particles}.
\newblock V$\overline{\mathrm{y}}$shaya shkola, Moscow, 1977.
\newblock (In Russian).

\bibitem{Physicedia-1}
{\em Physical encyclopedia, \textup{chief editor Prohorov~A.~M.}}, volume~1,
  Moscow, 1988. Sovetskaya Entsiklopediya (Soviet encyclopedia).
\newblock (In Russian).

\bibitem{bp-TM-general}
Bozhidar~Z. Iliev.
\newblock Transports along maps in fibre bundles.
\newblock JINR Communication E5-97-2, Dubna, 1997.
\newblock \\ LANL arXiv server, E-print No.\ dg-ga/9709016, 1997.

\bibitem{bp-normalF-LTP}
Bozhidar~Z. Iliev.
\newblock Normal frames and linear transports along paths in vector bundles.
\newblock \\ LANL arXiv server, E-print No.\ gr-qc/9809084, 1998.

\bibitem{bp-LTP-general}
Bozhidar~Z. Iliev.
\newblock Linear transports along paths in vector bundles. {I}.~{General}
  theory.
\newblock JINR Communication E5-93-239, Dubna, 1993.

\bibitem{bp-LTP-Cur+Tor-prop}
Bozhidar~Z. Iliev.
\newblock Linear transports along paths in vector bundles. {V}. {Properties} of
  curvature and torsion.
\newblock JINR Communication E5-97-1, Dubna, 1997.
\newblock \\ LANL arXiv server, E-print No.\ dg-ga/9709017, 1997.

\end{thebibliography}
\bibliographystyle{unsrt}

\addcontentsline{toc}{subsubsection}{This article ends at page}

\end{document}